# A theoretical investigation of the reaction between the amidogen, NH, and the ethyl, $C_2H_5$, radicals: a possible gas-phase formation route of interstellar and planetary ethanimine

Nadia Balucani[1,2,3], Dimitrios Skouteris,[4] Cecilia Ceccarelli,[2] Claudio Codella,[3] Stefano Falcinelli,[5] Marzio Rosi[5,6]
[1]*Dipartimento di Chimica, Biologia e Biotecnologie, Università degli Studi di Perugia, 06123 Perugia, Italy*
[2] *IPAG, Université Grenoble Alpes, 38000 Grenoble, France*
[3] *INAF, Osservatorio Astrofisico di Arcetri, 50125 Firenze, Italy*
[4] *Scuola Normale Superiore, 56126 Pisa, Italy*
[5] *Dipartimento di Ingegneria Civile e Ambientale, Università degli Studi di Perugia, 06125 Perugia, Italy*
[6] *CNR-ISTM, 06123 Perugia, Italy*

**Abstract**

The reaction between the amidogen, NH, radical and the ethyl, $C_2H_5$, radical has been investigated by performing electronic structure calculations of the underlying doublet potential energy surface. Rate coefficients and product branching ratios have also been estimated by combining capture and RRKM calculations. According to our results, the reaction is very fast, close to the gas-kinetics limit. However, the main product channel, with a yield of ca. 86-88% in the range of temperatures investigated, is the one leading to methanimine and the methyl radical. The channels leading to the two *E-, Z-* stereoisomers of ethanimine account only for *ca*. 5-7% each. The resulting ratio [*E*-$CH_3CHNH$]/[*Z*-$CH_3CHNH$] is *ca*. 1.2, that is a value rather lower than that determined in the Green Bank Telescope PRIMOS radio astronomy survey spectra of Sagittarius B2 North (*ca.* 3). Considering that ice chemistry would produce essentially only the most stable isomer, a possible conclusion is that the observed [*E*-$CH_3CHNH$]/[*Z*-$CH_3CHNH$] ratio is compatible with a combination of gas-phase and grain chemistry. More observational and laboratory data are needed to definitely address this issue.

## 1. Introduction

Imines are a family of N-containing organic compounds of great interest in the context of space prebiotic chemistry, because the small members of the family are simple enough to be formed in the extreme conditions of the interstellar medium, but they are very reactive and have the capability to evolve into much more complex species once brought to denser and chemically active environments (Woon, 2002; Balucani, 2009; Balucani, 2012; Loison et al., 2015). The strong propensity to react in dense media is caused by the presence of a carbon–nitrogen double bond which easily opens up providing an excellent addition site (Skouteris et al., 2015; Vazart et al. 2015). Having already an N-atom bound to a C-atom, small imines are believed to be valid precursors of small aminoacids like glycine/alanine or nucleobases (Woon, 2002). In particular, the smallest aldimine, namely methanimine ($CH_2$=NH), has been called into play to account for the formation of glycine by various mechanisms (Woon, 2002; Rimola et al, 2011; Rimola et al., 2012). Considering all these aspects, imines, together with the more familiar class of nitriles, might well represent the link between interstellar matter and the N-rich complex molecules from which life possibly emerged on primitive Earth (Caselli and Ceccarelli, 2012; Ehrenfreund et al., 2000; Chyba et al. 1990).

The first detection of methanimine dates back to 1973 (Godfrey et al., 1973), while its presence in several hot cores has been the subject of recent campaigns of detection (Suzuki et al. 2016; Widicus Weaver et al., 2017). In addition to methanimine, other imines have been recently identified in Green Bank Telescope PRIMOS radio astronomy survey spectra of Sagittarius B2 North (Sgr B2(N)), namely, ketenimine ($CH_2$=C=NH) (Lovas et al., 2006), cyanomethanimine (HN=CHCN) (Zaleski et al., 2013) and another aldimine, that is, ethanimine ($CH_3$CH=NH) (Loomis et al., 2013). Until recently, no detection of imines in low mass protostars has been reported and only upper limits have been given (Suzuki et al., 2016; Melosso et al., 2018). Ligterink et al. (2018) have finally identified methanimine towards IRAS 16293–2422B.

Remarkably, Loomis et al. (2013) detected ethanimine in both isomeric forms *E*- and *Z*-, with column densities of $7.00\times10^{13}$ cm$^{-2}$ and $2.3\times10^{13}$ cm$^{-2}$, respectively. Among the possible formation routes of ethanimine (see below for more details), Quan et al. (2016) have suggested the gas-phase reaction NH + $C_2H_5$, for which, however, there are no experimental data nor theoretical evaluations of the rate coefficient. Because of the great uncertainty associated to the NH + $C_2H_5$ reactive system, we have started a dedicated investigation by means of electronic structure calculations of the stationary points along the minimum energy path. Preliminary data have been reported in Rosi et al. (2018a). In this contribution, we have refined the electronic structure calculations of the transition states associated to all possible reaction channels. We have also performed a kinetic analysis and, given the complexity of this multichannel reaction, we have provided the product branching ratio at the temperatures of relevance for interstellar objects. In particular, we have carefully analyzed the formation routes of the two *E*-,*Z*- stereoisomers of ethanimine in an attempt to verify whether this gas-phase route is able to account for the observed *Z*-to-*E* isomer abundance ratio. Before presenting the results of our work, a brief summary of what has been suggested to account for the formation of ethanimine and other imines in the interstellar medium is presented to motivate the need of a reliable estimation of rate coefficients and product branching ratio for the title reaction.

### 1.1 Previous work on the formation of ethanimine and other related imines in the interstellar medium

While the formation of nitriles in extraterrestrial environments has been widely explored and their formation routes are well-established (see, for instance, Kaiser and Balucani, 2001; Balucani et al., 2000), more uncertain is the mechanism of formation of imines. Quan and Herbst (2007) suggested that ketenimine is likely formed by the dissociative recombination of protonated acetonitrile, $CH_3CNH^+$. Vazart et al. (2015) identified an easy gas-phase formation route for cyanomethanimine. A recent paper by Suzuki et al. (2016) reported the first comprehensive attempt to model methanimine abundance in hot cores. According to their rather complete gas-grain chemical simulations, hot core $CH_2NH$ is mostly formed in the gas phase by neutral–neutral reactions, rather than being the product of thermal evaporation from the surface of dust grains. Finally, Quan et al. (2016) recently attempted to model the abundance of ethanimine in Sgr B2(N) and concluded that $CH_3CHNH$ is mainly formed on grain surfaces by hydrogenation of $CH_3CN$, as previously suggested by Loomis et al. (2016) according to the scheme

$$CH_3CN_{(ice)} + H_{(ice)} \rightarrow CH_3CNH_{(ice)} \quad \text{(R1a)}$$

$$CH_3CNH_{(ice)} + H_{(ice)} \rightarrow CH_3CHNH_{(ice)} \quad \text{(R1b)}$$

However, the gas-grain network of reactions employed by Quan et al. is affected by incomplete information on the included reactions. First of all, even though hydrogenation of $CH_3CN$ on the icy mantles of interstellar grains appears to be a reasonable process, the formation of ethanimine, that is, the species resulting from the partial hydrogenation of acetonitrile, can be questioned if one considers the work by Theule et al. (2011), who have characterized a similar process involving HCN in laboratory experiments. In that work, methanimine, possibly resulting from partial hydrogenation of HCN, was never observed. Theule et al. (2011) explained the failure to observe $CH_2NH$ by suggesting that hydrogenation of methanimine is so favorable that it leads all the way to the fully hydrogenated counterpart, that is, methylamine ($CH_3NH_2$), even on small exposure to H atoms. In the same vein, it appears improbable that hydrogenation of $CH_3CN$ stops at the formation of ethanimine, rather than proceeding efficiently all the way to ethylamine according to

$$CH_3CHNH_{(ice)} + 2H_{(ice)} \rightarrow CH_3CH_2NH_{2(ice)} \quad \text{(R1c)}$$

Incidentally, while methylamine has been observed in interstellar objects (see Sleiman et al., 2018, and references therein), to the best of our knowledge there has been no detection of ethylamine.

Another process occurring on ice has been considered by Quan et al. (2016), that is, the radical-radical recombination reaction

$$CH_{3(ice)} + H_2CN_{(ice)} \rightarrow CH_3CHNH_{(ice)} \quad \text{(R2)}$$

Being a radical-radical recombination, the reaction has been included in the model by Quan et al. (2016) as a barrierless process. However, if one carefully considers the spin density of the $H_2CN$ radical (the unpaired electron is mostly localized on the N-atom), it appears quite obvious that the barrierless recombination process can only lead to the isomer $H_2C{=}N{-}CH_3$ (N-methyl methanimine), rather than to ethanimine ($CH_3{-}CH{=}NH$). $H_2C{=}N{-}CH_3$ is a stable closed-shell molecule, with an enthalpy of formation of only 39 kJ/mol higher than that of ethanimine, which is the most stable $C_2H_5N$ isomer. Therefore, only a severe rearrangement of the pre-existing chemical bonds can bring to ethanimine formation in reaction (R2) and this process requires surmounting a high energy barrier. For instance, according to the calculations by Balucani et al. (2010), the isomerization of the related radical species

$$CH_3NHCH_{2\,(g)} \rightarrow CH_3CH_2NH_{(g)} \quad \text{(R3)}$$

is characterized by a transition state located at +246.3 kJ/mol (see also below). And, as has been well illustrated by several works (Woon 2002; Rimola et al. 2014; Enrique-Romero et al., 2016, Rimola et al., 2018) the presence of ice molecules can only reduce a similar barrier height by a minimal part. In conclusion, if reaction (R2) is operative in interstellar ice, the final product will not be ethanimine, but its isomer N-methyl methanimine.

As already mentioned, both the *E*- and *Z*- isomers have been detected by Loomis et al. (2013). The *Z*-isomer is less stable than the *E*-isomer by 4.60 kJ/mol according to the calculations by Quan et al. (2016). This value has been refined to 2.77 kJ/mol in higher level calculations by Melli et al. (2018). The isomerization barrier is instead very high (115.8 kJ/mol according to the high level calculations of Melli et al.) because it is necessary to break the double bond of C=N to move from the *Z*- to *E*- isomers and *vice versa*. In this situation, it is not clear how ice-assisted chemistry can actually reproduce the observed difference, as tunneling through the high isomerization barrier can only favor the more stable isomer independently from the original amount formed of the two isomeric species. Were the grains involved in the ethanimine formation, only the *E*-isomer should be observed because the thermal population at the temperature of interstellar ice (10 K) favors the *E*- isomer by 14 orders of magnitude.

Among the gas-phase reactions considered by Quan et al. (2016), the reaction $NH+C_2H_5$ is certainly the best candidate, because the analogous reaction $NH+CH_3$ was seen to be the dominant formation route of methanimine in the model by Suzuki et al. (2016). In the absence of any information about the $NH+C_2H_5$ reaction, Quan et al. (2016) referred to the kinetic experiments by Stief et al. (1995) on the reaction $N+C_2H_5$ and employed a rate coefficient of $8.25\times10^{-12}$ cm$^3$ s$^{-1}$ for both $N+C_2H_5 \rightarrow$ *E*-$CH_3NCH$ + H and $NH+C_2H_5 \rightarrow$ *E*-$CH_3CHNH$ + H and of $2.75\times10^{-12}$ cm$^3$ s$^{-1}$ for both $N+C_2H_5 \rightarrow$ *Z*-$CH_3NCH$ + H and $NH+C_2H_5 \rightarrow$ *Z*-$CH_3CHNH$ + H (no T dependence was considered for these four reactions). However, *i)* atomic nitrogen is not isoelectronic with NH and, therefore, there is no reason to expect *a priori* a similar reactivity for the $N+C_2H_5$ and $NH+C_2H_5$ reactive systems; indeed, as we are going to see in this work, the $NH+C_2H_5$ reaction scheme is more complex than that of $N+C_2H_5$; *ii)* Stief et al. only observed the occurrence of the $C_2H_4$+NH and $H_2CN+CH_3$ channels and provided only an upper limit of <5% for all the other possible channels; *iii)* formation of $CH_3CHN$+H and not of $CH_3NCH$+H (indicated by Quan et al.) is feasible according to the calculations by Yang et al. (2005) on the $N+C_2H_5$ system; *iv)* it is not clear how the rate coefficients leading to *Z*- or *E*-isomers of ethanimine have been partitioned by Quan et al. (a ratio of 1:3 has been used, but no explanations are provided for such a choice).

In conclusion, a reliable estimate of the rate coefficient and product branching ratio for the title reaction is highly desirable. Given that an experimental determination is difficult to pursue because of the involvement of two radical species, we have applied a combination of electronic structure and kinetics calculations based on capture and statistical theories that are able to provide a realistic estimate, as already proved for reactions where experimental rate coefficients and/or product branching ratios are available (see, e.g., Sleiman et al., 2018; Leonori et al., 2009).

## 2. Computational Methodology for Electronic Structure and Kinetics Calculations

The whole reaction is assumed to take place exclusively in the lowest doublet electronic state of the [$C_2H_6N$] system. The global potential energy surface (PES) is a subset of the PES of the reaction $N(^2D)+C_2H_6$ that we have already characterized to assist the interpretation of the crossed molecular beam experiments of Balucani et al. (2010). However, since the total available energy is lower by ca. 140 kJ/mol, the system explores only specific regions of the global PES. In addition, the discrimination between the two *E-.Z-* isomers was not considered in previous calculations because it was not relevant for the results of the crossed molecular beam experiments.

We have calculated the stationary points employing a computational strategy which has already been utilized with success in several cases (see, for example, Rosi et al., 2018b; Sleiman et al., 2018; Skouteris et al. 2015; Leonori et al., 2013; Balucani et al., 2012; Balucani et al., 2009; de Petris et al., 2005). In particular, we optimize both energy minima and saddle points using density functional theory (DFT), making use of the B3LYP functional (Stephens et al., 1994; Becke, 1993). Subsequently, all stationary points (minima and saddle points) were subjected to energy refinement at the coupled cluster level, employing both single and double excitations and using a perturbative estimate of the effect of triple excitations (the CCSD(T) level, (Olsen et al., 1996; Raghavachari et al., 1989; Bartlett, 1981)). The correlation consistent aug-cc-pVTZ basis set has been used for both methods (Dunning, 1989). Saddle points (transition states) were located on the PES by making use of the synchronous transit-guided quasi-Newton method of Schlegel and coworkers (Peng et al., 1996; Peng and Schlegel, 1993). Vibrational frequencies were determined at the optimized geometries using the B3LYP/aug-cc-pVTZ method so that the nature of the stationary point was established (minimum if all frequencies are real and saddle point if exactly one frequency is imaginary). Transition states were connected with the corresponding reactant and product species through performing intrinsic reaction coordinate (IRC) calculations (Gonzalez and Schlegel, 1990; Gonzalez and Schlegel, 1989). The zero-point energy corrections, as calculated at the B3LYP/aug-cc-pVTZ level, were added to both B3LYP and CCSD(T) energies. The Gaussian 09 program suite (Frisch et al., 2009) was exclusively used for all calculations, while the vibrational frequencies were analyzed using Molekel (Portmann and Lüthi, 2000; Flükiger et al., 2000). Finally, some parts of the PES were computed also at the more accurate W1 level of theory (Martin & de Oliveira, 1999; Parthiban & Martin, 2001). We remind that in the W1 method the geometry optimization and the evaluation of the frequencies are performed at the B3LYP/VTZ + d level while the energies are computed at the CCSD(T)/AVDZ + 2d, CCSD(T)/AVTZ + 2d1f, CCSD/AVQZ + 2d1f level of theory (AVnZ is for aug-cc-pVnZ with n = D, T, Q). This method reproduced very nicely the experimental heat of formation of $CH_2NH$ and is expected to perform well for the system under study in this work (de Oliveira et al., 2001).

Subsequently, kinetics calculations were performed on the calculated PES, as previously done in Skouteris et al. (2018) and Skouteris et al. (2017) for similar reactive systems. Initially, the capture (Langevin) model is used to calculate the rate coefficient for the initial association of NH and $CH_3CH_2$. The long-range interaction potential (determined as a series of points through quantum calculations) was fitted to a $V(r) = -C/r^6$ equation (the typical interaction equation for two neutral species). Having obtained an association rate constant, we have calculated the corresponding dissociation rate coefficient through the detailed balance principle

$$K_{diss}(E) = k_{ass}(E) \times \rho_{reac}(E) / \rho_{comp}(E) \quad \text{(E1)}$$

where $k_{ass}(E)$ is the association rate coefficient (as a function of energy), $\rho_{reac}(E)$ is the density of states per unit volume of the associating reactants at energy E and $\rho_{comp}(E)$ is the density of states of the initial complex at energy E. Through this scheme, we have seen that the dissociation rate coefficient is essentially negligible (the large depth of the potential energy well raises the density of states of the complex). From this, we can deduce that the initial association rate coefficient equals the sum of all rate coefficients leading to specific products, i.e. the total reaction rate constant.

The microcanonical reaction rate coefficient for each subsequent unimolecular step is calculated using the Rice-Ramsperger-Kassel-Marcus (RRKM) scheme, whereby the rate coefficient is given by the expression

$$k(E) = \frac{N(E)}{h\rho(E)} \quad \text{(E2)}$$

where N(E) is the sum of states of the transition state, ρ(E) is the reactant density of states at energy E and h is Planck's constant. All densities of states are calculated through the convolution of the classical rotational density of states (and translational, in the case of the reactant channel) with the vibrational density of states. The vibrational density of states is taken to be the one deriving from the harmonic oscillator model and thus the Bayer-Swinehart algorithm is used for the convolution. The sum of states N(E) for each transition state is calculated through a convolution of the density of states and the energy-dependent tunneling probability (what would be a threshold function if tunneling were not included). Tunnelling is taken care of for each step by simulating each transition state with an Eckart barrier whose height is given by the relevant energy difference and its width calculated from the corresponding imaginary frequency.

Once all microcanonical rate coefficients have been calculated, a master equation is solved in order to determine rate coefficients for all product channels. After all microcanonical rate coefficients have been calculated, the master equation is solved for the particular energy in order to take account of the possibility of interconversion between intermediates. We set up a matrix $k$ of the rate constants such that the off-diagonal element $k_{ij}$ represents the rate constant from species j to species i and the diagonal elements are such that the sum of each column is 0. Moreover, we set up a concentration vector $c$ such that the element $c_j$ corresponds to the concentration of channel j. Then, all kinetics expressions can be written as a vector differential equation

$$\frac{dc}{dt} = kc \qquad \text{(E3)}$$

This is a linear differential equation and thus can be solved using standard methods. In order to determine the behavior of $c$ in the infinite future we diagonalise the matrix $k$ and determine its eigenvectors. These eigenvectors will either correspond to eigenvalues with a negative real part (vanishing in the infinite future) or 0 (stable eigenvectors). The initial concentration vector is written as a linear combination of all eigenvectors and, subsequently, those with a negative eigenvalue are discarded. What remains is the concentration vector in the infinite future, yielding the branching ratios of all channels. Finally, Boltzmann averaging is carried out over a range of temperatures to determine canonical (temperature-dependent) rate constants.

Finally, a comment on electron spin statistics is in order. The two reactants $CH_3CH_2$ and NH are in a spin-doublet and a spin-triplet state respectively, which implies a total of six spin substates. As a chemical bond must be formed between the reactants, the relevant two electrons (one from each species) to participate in the bond must be found in a singlet state. Among the six total spin states, two correspond to an overall doublet state and are reactive, while four correspond to an overall quartet state and are non-reactive. As a result, a statistical factor of 1/3 was applied to the total reaction rate constant in order to account for the non-reactive spin states of the reactants.

## 3. Results

### 3.1 Electronic structure calculations of minima along the minimum energy path

According to the present electronic structure calculations, there are eleven open channels for the title reaction:

$NH + C_2H_5 \rightarrow CH_3 + CH_2NH$ $\Delta H°_0$= -231(-243) kJ/mol (R4a)

$\rightarrow C_2H_4 + NH_2$ $\Delta H°_0$= -228(-240) kJ/mol (R4b)

$\rightarrow$ *E*-$CH_3CHNH$ + H $\Delta H°_0$= -203(-217) kJ/mol (R4c)

$\rightarrow$ *Z*-$CH_3CHNH$ + H $\Delta H°_0$= -200(-214) kJ/mol (R4d)

$\rightarrow CH_2CHNH_2$ + H $\Delta H°_0$= -186(-204) kJ/mol (R4e)

$\rightarrow CH_3NCH_2$ + H $\Delta H°_0$= -164(-178) kJ/mol (R4f)

$\rightarrow CH_2(NH)CH_2$ + H $\Delta H°_0$= -114(-130) kJ/mol (R4g)

$\rightarrow CHNH_2 + CH_3$ $\Delta H°_0$= -83(-94) kJ/mol (R4h)

$\rightarrow CH_3CNH_2$ + H $\Delta H°_0$= -59(-73) kJ/mol (R4i)

$\rightarrow CH_2NHCH_2$ + H $\Delta H°_0$= -28(-44) kJ/mol (R4j)

$\rightarrow CH_3N + CH_3$ $\Delta H°_0$= -14(-16) kJ/mol (R4k)

where the enthalpies of the reaction channels reported are those determined in the present work at the CCSD(T) and W1 (in parentheses) level of calculations. The complete potential energy surface has been derived at the CCSD(T)/aug-cc-pVTZ level; in Figure 1 the resulting minimum energy paths for the title reaction are shown (the channel (R4d) leading to the *Z*-$CH_3CHNH$ isomer has been omitted because of the figure congestion). Details on the pathways leading to both *E*-,*Z*-$CH_3CHNH$ isomer are shown in the close-up reported in Figure 2, where a different energy reference scale is used (see below). All enthalpy changes and reaction potential barrier heights associated with this reaction scheme are shown in Table 1, computed both using density functional theory (B3LYP/aug-cc-pVTZ level) and ab initio (CCSD(T)/aug-cc-pVTZ level) calculations. There is reasonably good agreement between the results obtained with the two methods. The quality of the present CCSD(T) calculations can be assessed by comparing the calculated enthalpy of reactions with the experimental values (when available). Considering the species involved in reaction (R4), such a comparison can be made for the channel (R4b). Using the values of the enthalpy of formation at 0 K and 298 K for the species NH, $C_2H_5$, $C_2H_4$ and $NH_2$ recommended by Burkholder et al. (2015), the experimental values are $\Delta H°_{0,exp}$= -238.7 ± 2.3 kJ/mol and $\Delta H°_{298,exp}$= - 239.3 ± 2.3 kJ/mol, to be compared with the values of the present determination of $\Delta H°_{0,CCSD(T)}$ = -228 kJ/mol and $\Delta H°_{298,CCSD(T)}$= - 230 kJ/mol. The comparison is not satisfactory and the disagreement is mainly due to the difficulty for the CCSD(T) method to well characterize the NH radical which is a triplet species and for which correlation effects are important. To have better values for the most relevant reactions channels, we have performed W1 calculations for selected stationary points. In this case, the experimental reaction enthalpies for (R4b) compare very nicely (within the experimental uncertainty) with the calculated values which are $\Delta H°_{0,W1}$ = -240 kJ/mol and $\Delta H°_{298,W1}$= - 241 kJ/mol (see Table 2).

3.1.1 The *E, Z*-$CH_3CHNH$ formation pathways

Since the reaction scheme is quite congested, we will first describe only the portion of the PES including the pathways leading to the two isomers *E, Z*-$CH_3CHNH$ (see Figures 1 and 2) and to $CH_2NH$.

The initial association adduct $CH_3CH_2NH$ is formed by the interaction of NH in the electronic ground state (triplet, $^3\Sigma^-$) with the ethyl ($C_2H_5$) radical. The adduct can be formed in two different geometries, depending on the relative position of the nitrogen lone pair and the N-H bond (*E*-$CH_3CH_2NH$ and *Z*-$CH_3CH_2NH$, see Figure 2). Both are energetically more stable than the separated reactants by 311.3 and 308.2 kJ/mol (at the CCSD(T) level of calculations), respectively, and there is no barrier associated with the two addition processes. The two isomers easily interconvert through a very small barrier located at an energy of -304.7 kJ/mol (at the CCSD(T) level of calculations) with respect to the reactants asymptote. Both of them can isomerize to $CH_3CHNH_2$ (the absolute minimum of the PES, located at – 339.8 kJ/mol, at the CCSD(T) level) via an H-shift from the central C atom to the N atom by overcoming a quite high barrier located at -164.2 and -165.0 kJ/mol with respect to the reactants asymptote. In addition, *E*-$CH_3CH_2NH$ can undergo a C-C bond fission producing $CH_3$+$CH_2NH$ or a C-H bond fission producing the *E*- isomer of ethanimine and atomic hydrogen. The same is true also for the *Z*- addition intermediate that produces either $CH_3$+$CH_2NH$ or *Z*-$CH_3CHNH$ + H. In both cases, the methyl loss channel is favored over the H-displacement as it requires to surmount a lower barrier locater at -197.8 kJ/mol, to be compared with -180.1 kJ/mol for *E*-$CH_3CH_2NH$$\rightarrow$ *E*-$CH_3CHNH$ + H or - 177.8 kJ/mol for *Z*-$CH_3CH_2NH$$\rightarrow$ *Z*-$CH_3CHNH$+ H (at the CCSD(T) level of calculations). We, therefore, expect that the channel (R4a) leading to methanimine + $CH_3$ will prevail over the channels (R4c)/(R4d).

The $CH_3CHNH_2$ intermediate can also produce *E*-,*Z*-$CH_3CHNH$ by undergoing a N-H bond fission. In addition, it can also form $CH_2CHNH_2$ + H in a slightly less exothermic channel (- 185.7 kJ/mol, at the CCSD(T) level) or $CHNH_2$ + $CH_3$ and $CH_3CNH_2$ + H in significantly less exothermic channels (- 82.7 kJ/mol and - 58.9 kJ/mol, respectively, at the CCSD(T) level). Starting from this intermediate, therefore, formation of *E*-,*Z*-$CH_3CHNH$ + H would be by far the dominant reaction channels.

Finally, according to the present CCSD(T)/aug-cc-pVTZ level) calculations, the energy gap between *E*-,*Z*-$CH_3CHNH$ is 2.9 kJ/mol, a value which compares very well with that determined by Melli et al. at the CCSD(T)/CBS+CV level corrected for the anharmonic ZPE at the B2PLYP-D3BJ/maug-cc-pVTZ-dH level (2.77 kJ/mol). The *E*-→*Z*- interconversion barrier is found to be 117.9 kJ/mol, again in excellent agreement with the calculations by Melli et al. (2018).

As we are going to see in Sec. 3.2, not only this portion of the PES is the one of interest to characterize the formation of aldimines, but is also the part experienced by most of the reactive flux. For this reason, given the time-consuming nature of W1 calculations, we have limited them to this portion of the PES. In Figure 2, we have reported this part of the PES scheme with an indication of the W1 and CCSD(T) energy values. Once the first intermediate is formed, the relative values of the stationary points are the ones really controlling the rate coefficients for each individual step. For this reason, in Figure 2 we have reported the energy values by referring to the energy content of the first addition intermediate *E*-$CH_3CH_2NH$ as the reference value (in this way it is much easier to compare the two sets of values without the effect of the CCSD(T) problems in treating the NH radical). As is well visible by inspecting the relative values of the energy of the stationary points, W1 and CCSD(T) nicely agree (all of them are within 5 kJ/mol with W1 values systematically lower). Therefore, we can confirm that the significant deviation in the enthalpies of reactions is mostly due to the poor characterization of the NH reactant in the CCSD(T) calculations.

### 3.1.2 The rest of the [$C_2H_6N$] PES

Once the $CH_3CHNH_2$ intermediate is formed, it can undergo further isomerization through another migration of an H atom to the $CH_2CH_2NH_2$ species. This latter one is less stable by 38.9 kJ/mol and its formation requires overcoming a barrier located at -146.2 kJ/mol with respect to the reactant asymptote. Once $CH_2CH_2NH_2$ is formed, it can dissociate to ethylene and $NH_2$ (this channel is exothermic by -228.3 kJ/mol) through a reaction barrier located at -207.1 kJ/mol. Alternatively, $CH_2CH_2NH_2$ can eliminate an H atom giving rise to $CH_2CHNH_2$ + H (this channel is exothermic by 185.7 kJ/mol with respect to the reactants); this reaction, however, requires the overcoming of a barrier of 140 kJ/mol. The initial association adduct $CH_3CH_2NH$ can also isomerize to $CH_3NHCH_2$ which is only 7.3 kJ/mol less stable; this reaction requires an energy as high as 253.6 kJ/mol to overcome the barrier height, which is still, however, below the reactants. $CH_3NHCH_2$, once formed, can dissociate to $CH_3$ + $CH_2NH$, which is energetically higher by 73.5 kJ/mol which shows a barrier of 125.3 kJ/mol. $CH_3NHCH_2$ shows also other dissociation channels, which are however less stable. In particular, it can eliminate an H atom giving rise to $CH_3NCH_2$ + H ($\Delta H°_0$ = - 163.7 kJ/mol; barrier height almost equal to the product energy), $CH_2NHCH_2$ + H ($\Delta H°_{0=-}$ 27.5 kJ/mol; barrier height 276.8 kJ/mol) or the cyclic species $CH_2(NH)CH_2$ + H ($\Delta H°_0$ -114.2 kJ/mol; barrier height 268.8 kJ/mol). $CH_3NHCH_2$, through an H migration, can also isomerize to $CH_3NCH_3$; this reaction shows a high barrier ($\Delta H°_0$ 2.8 kJ/mol; barrier height 172.3 kJ/mol). $CH_3NCH_3$, once formed, can dissociate to $CH_3N$ + $CH_3$; the products are much less stable than $CH_3NCH_3$ (by 287.0 kJ/mol) but this exit channel is still below the reactants. Alternatively, it can eliminate an H atom forming the already mentioned $CH_3NCH_2$ + H.

### 3.2 Rate coefficients and product branching ratio

The RRKM analysis of the complete PES calculated at the CCSD(T) level revealed that only 3 out of the 11 possible channels actually contribute to the global rate coefficients. The channels are, in order of decreasing importance, those leading to methanimine (R4a), *E*-ethanimine (R4c), and *Z*-ethanimine (R4d). The contribution of all other channels is negligible, either because those channels are characterized by the presence of much higher transition states or because too many rearrangements are necessary to reach the final configuration (this is, for instance, the case of the channel leading to $C_2H_4$ + $NH_2$).

After having established than only channels (R4a), (R4b) and (R4c) have a significant yield when considering the entire PES, we have repeated the kinetics calculations using the W1 energies for the reduced PES. In Fig. 3 are shown the rate coefficients as a function of the temperature for the three dominant reaction channels. Since the entrance channel is barrierless, the global rate coefficient is very high at the gas kinetics limit; once partitioned according to the branching ratio, however, only the dominant channel remains in the $10^{-10}$ range, while the rate coefficients of the two channels leading to the *E-,Z*-ethanimine isomers do not exceed the $10^{-11}$ range. Moreover, it can be seen from Fig. 2 that, as expected from a capture-like rate coefficient, at low temperatures they rise steeply and afterwards they reach a plateau at higher temperatures. The rate constant has been fitted to a modified Arrhenius law

$$k=\alpha \times (T_{gas}/300\ K)^{\beta} \times \exp[-\gamma/T_{gas}] \qquad (E4)$$

which essentially implies a linear variation of the activation energy with the temperature

$$E_a / K = \gamma + \beta\, T \qquad (E5)$$

The best-fit resulting coefficients valid in the T range between 10 and 300 K are reported in Table 2, while the branching ratios at selected temperatures representative of interstellar objects are shown in Table 3. As visible, at all the temperatures considered the channel leading to methanimine + $CH_3$ is by far the dominant one accounting for more than 85% of the global yield at all temperatures considered. Also, the branching ratio between the *E*- and *Z*- isomers of ethanimine are comparable, being *ca.* 1.1 in the entire analyzed range. Even though the *E*-isomer is slightly more abundant, the branching ratio is far from the factor of *ca*. 3 derived from the observation of Loomis et al. (2013).

The fact that the rate coefficients resulting from the W1 calculations (where the energies of the relevant intermediates and TS differ from the CCSD(T) ones typically by 3-5 kJ/mol) are essentially identical to the ones resulting from CCSD(T) confirms that the rate coefficients are reasonably robust with respect to quantum chemistry errors.

## 4. Discussion and astrophysical implications

A first comment we would like to make is that the NH + $C_2H_5$ reaction appears to be generally more complex than the reaction N + $C_2H_5$ which is characterized by only seven open channels according to the calculations of Yang et al. (2005). Unfortunately, Yang et al. did not perform a kinetic analysis of their PES so that they could not establish RRKM branching ratios for the reaction N + $C_2H_5$. However, we can note some important differences. For instance, the channel leading to $C_2H_4$ + NH only needs two isomerizations from the initial addition intermediate to reach the right configuration for the NH elimination, while in the analogous process for the title reaction three isomerizations are necessary before reaching the $CH_2CH_2NH_2$ intermediate, that is the only one that can release an $NH_2$ radical. In addition, only five H-displacement channels exist for the N+$C_2H_5$ reaction, while seven H-displacement channels characterized NH+ $C_2H_5$. Finally, at 298 K the estimated rate coefficient of the title reactions is almost twice that determined by Stief et al for N+$C_2H_5$ and the H-displacement channels were found to be negligible, while in the present case they account for 13-14% of the global reaction.

Another point of interest is to compare the rate coefficients that we have determined theoretically with those employed by Quan et al. (2016) in their model, that is, $8.25\times10^{-12}$ $cm^3$ $s^{-1}$ for NH+$C_2H_5\rightarrow$*E*-$CH_3$CHNH + H (R4c) and of $2.75\times10^{-12}$ $cm^3$ $s^{-1}$ for NH+$C_2H_5\rightarrow$*Z*-$CH_3$CHNH + H (R4d) with no temperature dependence. The estimated value for *E*-$CH_3$CHNH is not far from the one we have determined here: considering the T range of the model by Quan et al., $k_{(R4c)}$ varies from $8.2\times10^{-12}$ at 10 K up to $1.3\times10^{-11}$ at 200 K. However, the rate coefficient for *Z*-$CH_3$CHNH is larger by a factor of 2.6-4 being $7.2\times10^{-12}$ at 10 K and $1.1\times10^{-11}$ at 200 K. In other words, there is not a large difference between $k_{(R4c)}$ and $k_{(R4d)}$, as instead assumed by Quan et al. (2016). In addition, Quan et al. (2016) totally omitted in their model to consider the main channel (R4a) that will consume a large fraction of the reactants, thus reducing the total production rate of ethanimine.

A more general comment concerns the comparison between the *E-,Z-* isomers branching ratio that we have determined here and the relative abundance observed in Sgr B2(N). The NH+$C_2H_5$ reaction is certainly the best candidate to explain ethanimine formation in the gas phase. We can affirm this by considering the related work on methanimine formation by Suzuki et al. (2016). In their model, methanimine is formed in the gas-phase essentially via NH+$CH_3$, because during the cold collapsing phase methanimine formed in the ice is completely hydrogenated to methylamine. Also, in their model the saturated complex molecules like $NH_3$, $CH_4$, $CH_3OH$, and $CH_3NH_2$ formed on the dust sublimate during the warm-up phase and then generate the precursor NH and $CH_3$ radicals through gas-phase processes. Notably, methanimine does not have stereoisomers like ethanimine. Therefore, provided that some ethane is also sublimated from dust together with methane and the other saturated compounds, the distribution of the two isomers of ethanimine can give additional general information on aldimine formation routes in interstellar objects, as it is a very sensitive test for astrochemical models. We have already commented on the fact that, were ethanimine formed on the ice in the cold phase, irrespectively from the detailed mechanism of formation, tunneling through the very high isomerization barrier would inevitably favor the most stable isomer *E*-$CH_3$CHNH if only thermal sublimation is considered. The thermal population at possible ice temperatures provides a ratio [*E*-$CH_3$CHNH]/[*Z*-$CH_3$CHNH] of $6\times10^{14}$ at 10 K or $8\times10^{4}$ at 30 K.

However, were the gas-phase reaction NH+C2H5 the only formation mechanism of ethanimine, the [*E*-CH3CHNH]/[*Z*-CH3CHNH] ratio would be close to unity, namely approximately consistent with the one determined by Loomis et al. (2013). To be noted that, because of the very high isomerization barrier, once released in the gas-phase only chemical reactions could alter the *E*-/*Z*- ratio. Unfortunately, Quan et al. (2016) did not explain how they quantified the amount of the two isomers produced in their model. In all cases, a possible conclusion is that, if no other, yet unexplored, formation routes of ethanimine are actually contributing, the observed ethanimine is mostly produced in the gas-phase (where the process is controlled by kinetics and provides almost identical populations of the two isomers), with a much smaller contribution from the grain-surface (where the thermal population of the ethanimine adsorbed on ice would be E-/Z- about 10^5 at 30 K). Alternatively, non-thermal ice chemistry could be at work (see, for instance, Frigge et al. 2018 and Bergner et al. 2017).

More constraints on the observation and additional experimental or theoretical data on the possible formation routes of the two isomers (such as the determination of the population of the two isomers in hydrogenation experiments of acetonitrile) are strongly needed to finally solve the puzzle.

The results obtained for the title reaction are also of interest in the chemistry of the atmosphere of Titan, the massive moon of Saturn, which is characterized by a large fraction of nitrogen and a small percentage of methane and higher hydrocarbons. The formation routes of imines, amines, and nitriles in the context of Titan have been recently addressed by Loison et al. (2015). From an uncertainty propagation study and a global sensitivity analysis performed with the method developed by Hébrard et al. (2009), Loison et al. (2015) have determined a list of the key reactions which are responsible for the uncertainties on the mole fraction profiles calculated in their model. According to the recommendation of Loison et al. (2015), these reactions should be characterized from a chemical point of view to improve photochemical models of Titan's atmosphere. Reaction (4) belongs to the list of reactions which are critical to account for the $NH_3$ amount in the 100 km altitude region. It is included in their model considering only the occurrence of channel (R4b) with an estimated $k=1\times10^{-11}$ $cm^3$ $s^{-1}$ and channels (R4c)+(R4d) with an estimated $k=6.0\times10^{-11}(T/300)^{0.17}$ $cm^3$ $s^{-1}$. Therefore, the main channel (R4a) is missing, $k_{(R4b)}$ is largely overestimated (according to our determination its value is in the $10^{-14}$ $cm^3$ $s^{-1}$ range for the temperatures of interest for the upper atmosphere of Titan) while the value of rate coefficient leading to both isomers of ethanimine is not far (within a factor 2) from the present determination. In the light of the present work, the importance of the title reaction in the $NH_3$ budget of Titan needs to be reconsidered.

**Acknowledgments**

This work has been supported by MIUR "PRIN 2015" funds, project "STARS in the CAOS (Simulation Tools for Astrochemical Reactivity and Spectroscopy in the Cyberinfrastructure for Astrochemical Organic Species)", Grant Number 2015F59J3R. DS wishes to thank the Italian Ministero dell'Istruzione, Università e Ricerca (MIUR_FFABR17_SKOUTERIS) and the Scuola Normale Superiore (SNS_RB_SKOUTERIS) for financial support. Partially supported also by the European Research Council (ERC) under the European Union's Horizon 2020 research and innovation pro-gramme, for the Project "The Dawn of Organic Chemistry" (DOC), grant agreement No 741002 and by PRIN-INAF 2016 The Cradle of Life - GENESIS-SKA (General Conditions in Early Planetary Systems for the rise of life with SKA).

Table 1. Enthalpy changes (kJ/mol, 0 K) computed at the B3LYP/aug-cc-pVTZ and CCSD(T)/aug-cc-pVTZ levels of theory for selected reactions of the systems NH + $CH_3CH_2$. For comparison purposes, W1 energies are reported in parentheses for selected processes.

| | $\Delta H^0_0$ | | Barrier height | |
|---|---|---|---|---|
| | B3LYP | CCSD(T) | B3LYP | CCSD(T) |
| **NH ($^3\Sigma^-$) + $CH_3CH_2$ → $CH_3CH_2NH$ (*E*)** | -309 | -311 (-323) | | |
| **NH ($^3\Sigma^-$) + $CH_3CH_2$ → $CH_3CH_2NH$ (*Z*)** | -305 | -308 (-320) | | |
| **$CH_3CH_2NH$ (*E*) → $CH_3CH_2NH$ (*Z*)** | 4 | 3 (3) | 8 | 7 (7) |
| **$CH_3CHNH$ (*E*) → $CH_3CHNH$ (*Z*)** | 3 | 3 (3) | 107 | 118 (113) |
| **$CH_3CH_2NH$ (*E*) → $CH_3CHNH_2$** | -35 | -29 (-34) | 145 | 147 |
| **$CH_3CH_2NH$ (*Z*) → $CH_3CHNH_2$** | -39 | -32 (-37) | 140 | 143 |
| **$CH_3CHNH_2$ → $CH_2CH_2NH_2$** | 47 | 39 | 191 | 194 |
| **$CH_3CH_2NH$ (*E*) → $CH_3CHNH$ (*E*) + H** | 115 | 109 (106) | 131 | 131 (126) |
| **$CH_3CH_2NH$ (*Z*) → $CH_3CHNH$ (*Z*) + H** | 114 | 109 (106) | 130 | 131 (126) |
| **$CH_3CH_2NH$ (*E*) → $CH_3$ + $CH_2NH$** | 73 | 81 (80) | 105 | 114 (112) |
| **$CH_3CH_2NH$ (*Z*) → $CH_3$ + $CH_2NH$** | 69 | 78 (77) | 101 | 111 (109) |
| **$CH_3CHNH_2$ → $CH_3CHNH$ (*E*) + H** | 149 | 137 (140) | 152 | 148 |
| **$CH_3CHNH_2$ → $CH_3CHNH$ (*Z*) + H** | 153 | 140 (143) | 156 | 152 |
| **$CH_3CHNH_2$ → $CH_2CHNH_2$ +H** | 158 | 154 (153) | 159 | 154 |
| **$CH_3CHNH_2$ → $CH_3$ + $CHNH_2$** | 251 | 257 (263) | 255 | 259 |
| **$CH_3CHNH_2$ → $CH_3CNH_2$ + H** | 287 | 281 (284) | 287 | 281 |
| **$CH_2CH_2NH_2$ → $C_2H_4$ + $NH_2$** | 64 | 73 | 83 | 94 |
| **$CH_2CH_2NH_2$ → $CH_2CHNH_2$ + H** | 112 | 115 | 126 | 140 |

Table 2. Rate coefficients for the reaction channels (R4a), (R4c), and (R4d) that make a significant contribution to the title reaction. α, β, and γ are the coefficients that allow accounting for the temperature dependence of the rate coefficients according to the equation $k=\alpha \times (T_{gas}/300\ K)^{\beta} \times \exp[-\gamma/T_{gas}]$. Valid in the T range between 10 and 300 K.

| **Reaction** | **α ($cm^3\ s^{-1}$)** | **β** | **γ** |
|---|---|---|---|
| $NH + C_2H_5 \rightarrow CH_3 + CH_2NH$ (R4a) | $1.69 \times 10^{-10}$ | 0.180 | 0.490 |
| $NH + C_2H_5 \rightarrow$ *E*-$CH_3CHNH$ + H (R4c) | $1.34 \times 10^{-11}$ | 0.119 | 0.669 |
| $NH + C_2H_5 \rightarrow$ *Z*-$CH_3CHNH$ + H (R4d) | $1.19 \times 10^{-12}$ | 0.122 | 0.726 |

Table 3. Product branching ratios at selected temperatures.

| **Channel** | **BR @ T=10 K** | **BR @ T=100 K** | **BR @ T=300 K** |
|---|---|---|---|
| $CH_3 + CH_2NH$ (R4a) | 85.0 % | 86.1 % | 87.0 % |
| *E*-$CH_3CHNH$ + H (R4c) | 8.0 % | 7.4 % | 6.9 % |
| *Z*-$CH_3CHNH$ + H (R4d) | 7.0 % | 6.5 % | 6.1 % |

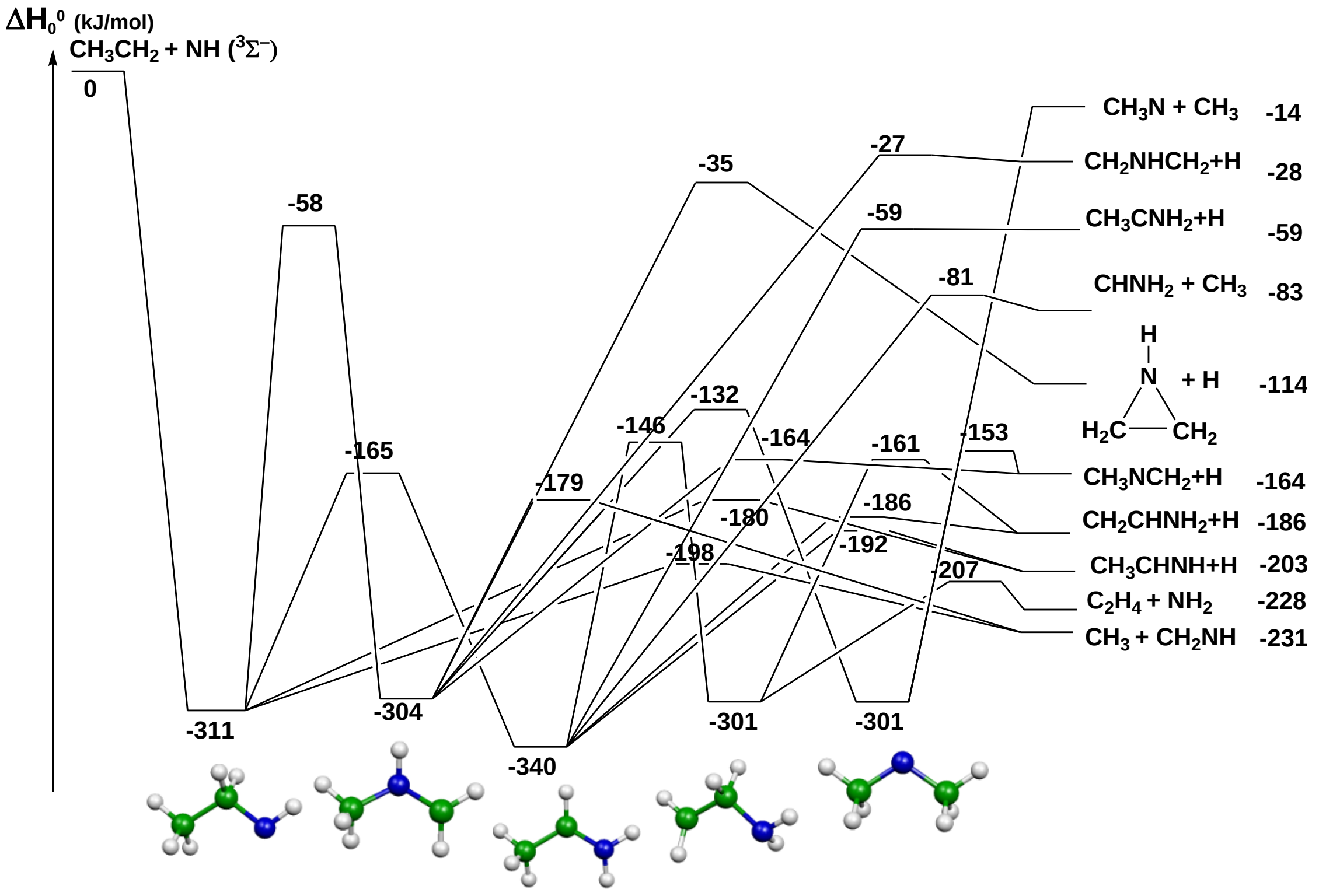

Figure 1: Schematic representation of the $NH+C_2H_5$ potential energy surface optimized at the CCSD(T) level. For simplicity, only the CCSD(T) relative energies (kJ/mol) are reported. B3LYP values are reported in Table 1. Only the channel leading to the *E*-isomer of $CH_3CHNH$ is shown (see text).

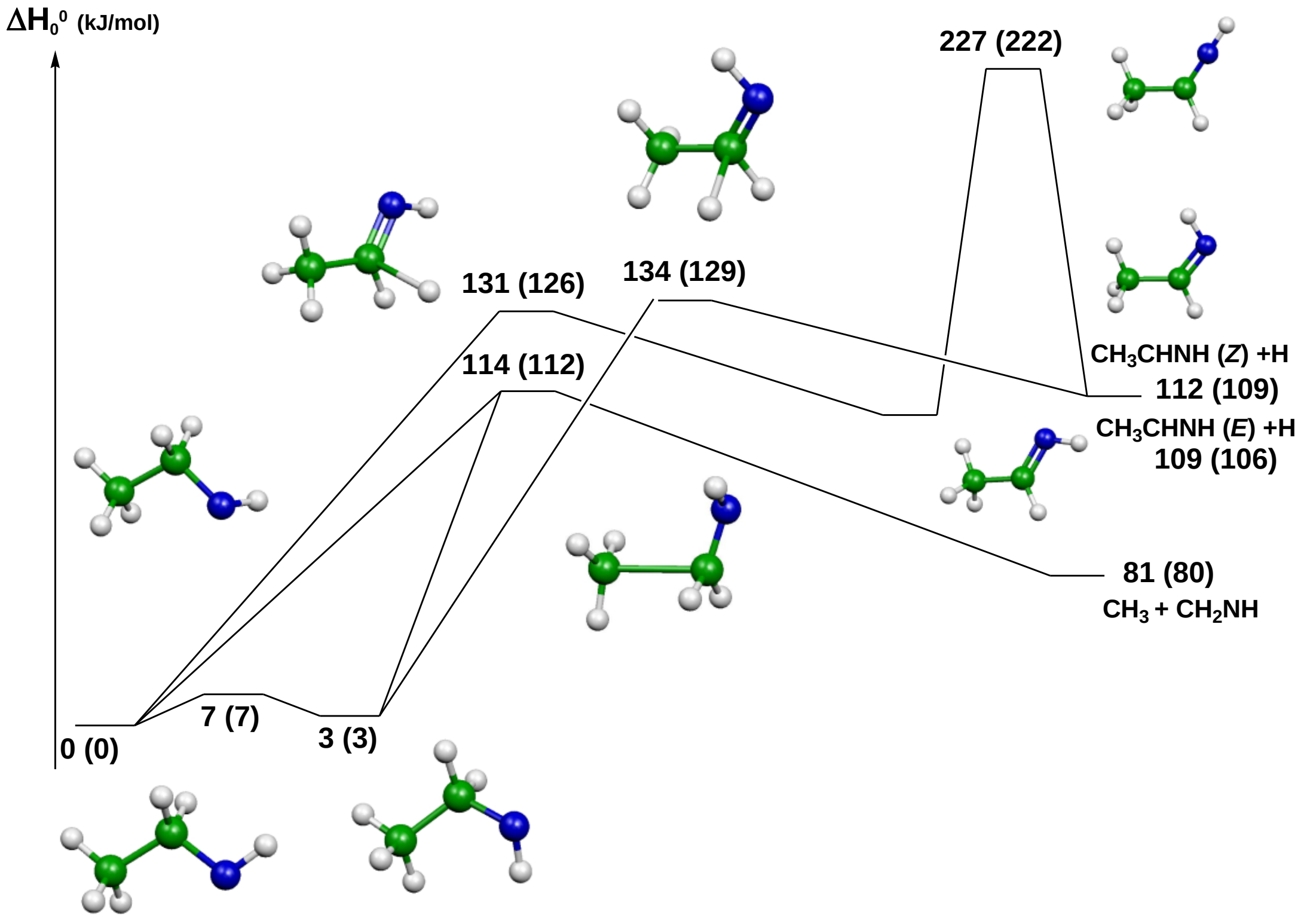

Figure 2: A close-up of the portion of the NH+$C_2H_5$ potential energy surface illustrating the possible pathways towards the formation of *E-,Z*-ethanimine. CCSD(T) and W1 (in parentheses) relative energies (kJ/mol) are reported with respect to the energy of the initial association intermediate *E*-$CH_3CH_2NH$ taken as zero.

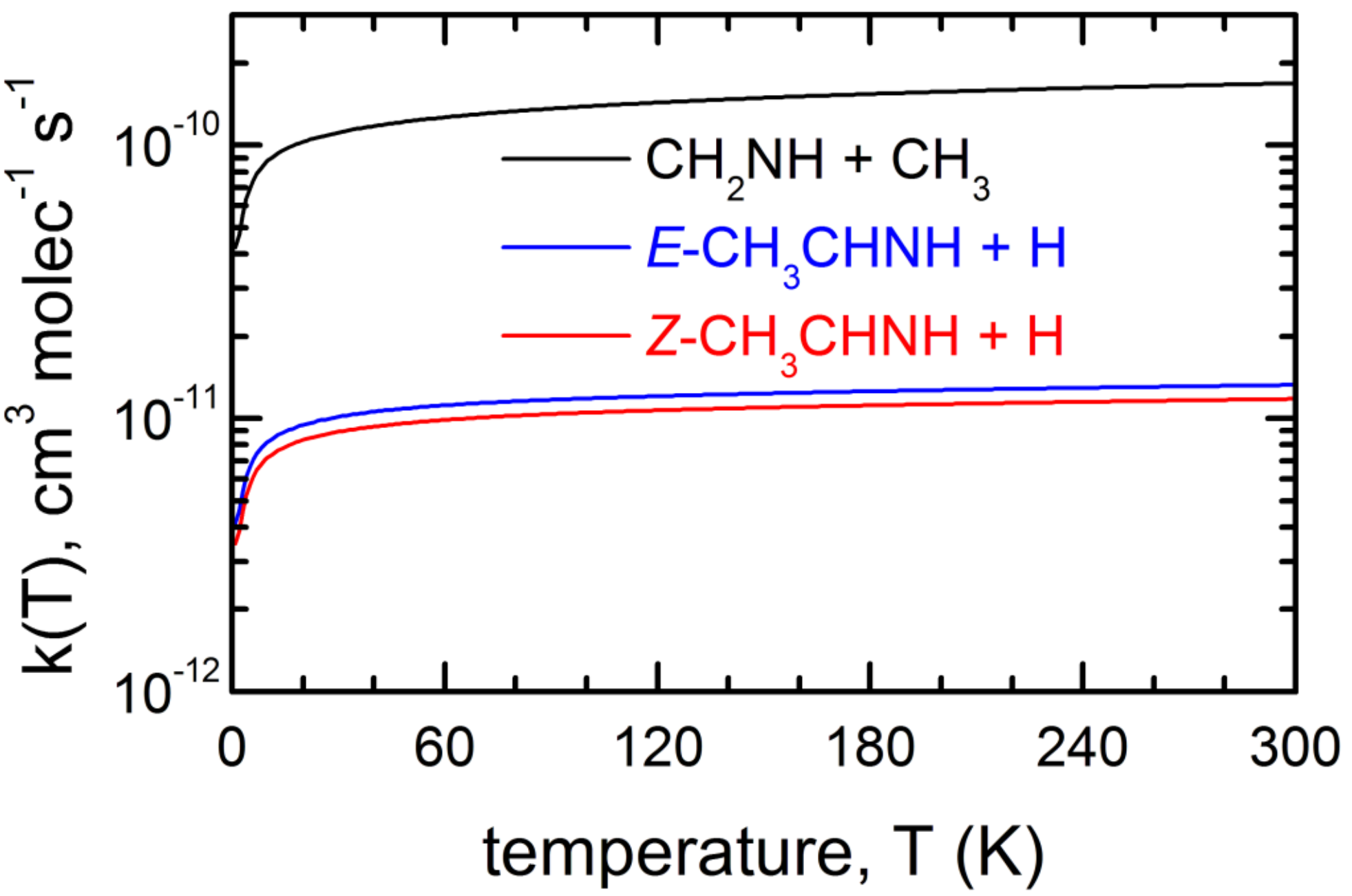

Figure 3: Rate coefficients as a function of temperature for the three channels with a significant yield.